\documentstyle[12pt]{article}
\def\PRL#1#2#3{{\sl Phys. Rev. Lett.} {\bf#1} (#2) #3}

\def\RMP#1#2#3{{\sl Rev. Mod. Phys.} {\bf #1} (#2) #3}

\def\JETP#1#2#3{{\sl Sov. Phys. JETP} {\bf #1} (#2) #3}

\newcommand{\nn}{\nonumber\\}\newcommand{\p}[1]{(\ref{#1})}
\begin{document}
\renewcommand{\thefootnote}{\fnsymbol{footnote}}
\thispagestyle{empty}
\begin{flushright}
{Preprint DFPD/93/TH/45\\
 June 1993}
\end{flushright}
\vspace{0.2cm}
\begin{center}
{\large\bf Helicity in Classical Electrodynamics and Turbulence\\}
\vspace{1cm}
Aleksej V. Chechkin,\footnote{e-mail:
stc\%stcep.kharkov.ua@relay.ussr.eu.net}
\vspace{0.5cm}\\
{\it Scientific and Technological Centre of Electrophysics\\
 Ukrainian Academy of Sciences, P/O Box 9410, 3rd Valter St.,
Kharkov,\\ 310108 the Ukraine}
\renewcommand{\thefootnote}{\dag}
\vspace{0.5cm}\\
Dmitrij P. Sorokin\footnote{Supported in part by the European Community
under contract CEE-SCI-CT92-0789}
\renewcommand{\thefootnote}{\ddagger}
\footnote{Permanent address: Kharkov Institute of Physics and
Technology, Kharkov, 310108, the Ukraine
e-mail address:
kfti\%kfti.kharkov.ua@relay.ussr.eu.net}\\
\vspace{0.5cm}
Dipartimento di Fisica ``G. Galilei'', Universit\'a di Padova\\

\smallskip
Instituto Nazionale di Fisica Nucleare, Sezione di Padova\\
Via Marzolo 8, 35131 Padova, Italia\\
\vspace{0.5cm} and\\
\vspace{0.5cm}
Vladimir V. Yanovsky$^*$
\vspace{0.5cm}\\
{\it Scientific and Technological Centre of Electrophysics\\
 Ukrainian Academy of Sciences, P/O Box 9410, 3rd Valter St.,
Kharkov,\\
310108 the Ukraine}

\vspace{1.cm}
{\bf Abstract}
\end{center}
\bigskip
For the electromagnetic fields, hydrodynamic media and
turbulent flows we consider the relationship between a topological
characteristic of vector fields known as helicity and the angular
momentum of the medium, and discuss, in this respect, the problem of
helicity and angular momentum transfer from  the electromagnetic field
to a medium.

\setcounter{page}0
\renewcommand{\thefootnote}{\arabic{footnote}}
\setcounter{footnote}0
\newpage

\section{Introduction}

Taking into account topology is one of the most important aspects in
studying various physical objects. Nontrivial topological
properties guaranty the stability of classical and quantum states of
a physical system, cause the possibility of existing in nature such
stable point-like objects as vortices, solitons, monopoles and
instantons, and may be a cause and explanation of several physically
observed phenomena (Abricosov vortices, Aharonov--Bohm effect, quantum
Hall effect etc.) (\cite{raj,vk} and references therein). The
topological characteristics of the system are topological invariants
conserved under continuous deformations of system parameters. From
physical point of view these topological invariants should correspond
to physical observables, i.e.  integrals of motion or conserved
charges, characterizing the dynamics of the system.  In this respect
establishing the relationship between topological and physical
characteristics has proved to be important for deeper understanding of
physical properties and the evolution of the system.

Among the topological invariants having been engaged in various fields
of modern physics the Hopf invariant \cite{h} takes one of the first
places. In 3--dimensional space it characterizes, in particular,  the
linking and the knot number of the integral lines of a vector field
$A_{\alpha}(x)$ ($\alpha,\beta=1,2,3$) describing a physical system.
The helicity of $A_{\alpha}(x)$, determined by the
integral
\begin{equation}
h={1\over
4\pi
c}\int{d^{3}x\varepsilon^{\alpha\beta\gamma}A_{\alpha}\partial_{\beta}
A_{\gamma}}
\end{equation}
 (where $\varepsilon^{\alpha\beta\gamma}$ is the unit
skew-symmetric tensor with $\varepsilon^{123}=1$), has the direct
relation to the Hopf invariant \cite{h}. The integral (1) does not
depend on the (local) metric properties of the 3--dimensional space
and, hence, characterizes global properties of the space and the vector
field.

 The properties of helical fields have been studied in detail, for
example, in the theory of magnetic dynamo \cite{m} and hydrodynamics
\cite{ty}. For instance, it is well known that for being able to
generate a large-scale magnetic field by a small-scale motion of
conducting fluid it is necessary for the turbulence to possess
helicity. One more example of helicity manifestation is the generation
of large-scale structures in hydrodynamic turbulence \cite{mst}. If a
small-scale turbulence of a fluid or gas possess helicity, then
large-scale instability leading to the generation of the vortex
structures arises.  The generation of the large-scale structures is
caused by the mirror noninvariance of the turbulence.

In plasmas the problem of current maintenance stimulated great
interest in the possibility of generating the current by injecting
helical electromagnetic waves \cite{t}, the problem being mainly
considered in a way similar to the dynamo effect and examined with
magneto-hydrodynamic (MHD) equations.

In quantum field theory formulated in $D=2+1$
space-time Eq.~(1), called the Chern--Simons term (or action), is the
basis for developing so called topological quantum field theories
\cite{s}. Being invariant under (abelian) gauge transformations of the
vector field ($A_{\alpha}(x)\rightarrow
A_{\alpha}(x)+\partial_{\alpha}\varphi(x)$, where $\varphi(x)$ is an
arbitrary function which at the infinity tends quickly enough to zero)
the Chern--Simons term allows one in $D=2+1$ to consider massive vector
fields without breaking the gauge invariance of the theory \cite{djt}.
An approach to the description of particles with fractional statistics
and spin \cite{lm}, which in their turn may occur to be relevant to
the problem of high--$T_c$ superconductivity \cite{a}, is also based on
the interaction of matter fields with gauge fields described by the
Chern--Simons action (1) (see \cite{w} and references therein).

In view of the variety of ways the helicity
manifests itself in physics and for deeper understanding of this
feature it seems of interest to study the meaning of the invariant (1)
for such well-studied physical objects as electromagnetic fields in
4-dimensional space-time, hydrodynamic media and turbulence, and to
find the connection of $h$ with observable physical characteristics of
the system, thus establishing the role of conservation laws in helicity
transfer from the field to the medium. Since the helicity characterizes
the `twisting' of the vector fields it is natural to assume that the
helicity is connected with the angular momentum of the system. For
$D=2+1$ Chern-Simons field theory this is well understood, but for
ordinary electromagnetic fields and turbulent flows, as
we are aware, these problems have not  been discussed in detail yet.

In the present paper we consider the relationship between the helicity
(1) and the angular momentum of the electromagnetic field, hydrodynamic
and turbulent flows, and discuss the problems of angular momentum and
helicity transfer from the electromagnetic field to the medium. It is
shown (Sections 2, 3) that the trace of the tensor of the spin momentum
flow density of the electromagnetic field is equivalent to the helicity
density of the electromagnetic field. Physically it means that helicity
density characterizes the value of the spin component of the angular
momentum flow passing per unit of time through the unit surface normal
to the direction of electromagnetic field propagation (characterized by
the Poyting vector) in the observation point. As is well-known, the
projection of the angular momentum onto the momentum of the
electromagnetic field is an invariant of the Puancar\'e group called
 the helicity (not to be confused with (1)!) of the photon.
To distinguish these two helicities we denote the
first of them by $h$ and the second one \hbox{by $s$}. In classical
electrodynamics $s$--helicity  corresponds to the degree of
electromagnetic wave polarization. As we shall see,
Eq.~(1) determines the degree of the circular
polarization of the electromagnetic wave and is proportional to the
mean value of angular momentum projection on to the Poynting vector.
 In the case of the plane wave Eq.(1) indeed coincides
with the expression for the classical analogue of the photon helicity
$s$, and, therefore, is conserved (physically observable) quantity.

In Section 4 by analogy with the electromagnetic case we consider the
relationship between the helicity and angular momentum for turbulent
flows, and introduce the kinetic definition of the helicity of particle
systems. This kinetic definition is similar to the hydrodynamic one,
but it allows us to get deeper insight into the microscopic nature of
the $h$--helicity.

In Section 5 a problem of angular momentum and
helicity transfer from the field to the medium is discussed by use of a
simple example of a charged particle propagating in a circular
polarized plane electromagnetic wave. It is shown that the particle can
acquire angular momentum from electromagnetic waves with nonzero
helicity.

\section{The angular momentum of the
electromagnetic field} A density tensor of the angular momentum of the
electromagnetic field may be constructed by the analogy with the
orbital momentum tensor of the classical particle \cite{ll}:
\begin{equation}\label{s}
M_{ij}^{ k}=x_{i}T_{j}^{ k}-x_{j}T_{i}^{ k},
\end{equation}
where
\begin{equation}\label{em} T_{ij}={1\over 4\pi
c}(F_{i}^{ k}F_{kj}-{1\over 4}g_{ij}F_{kl}F^{kl})
\end{equation}
is the
symmetric energy-momentum tensor and $F_{ik}=\partial_{i}A_{k}-
\partial_{i}A_{k}$ is the electromagnetic field stress tensor, $g_{ik}$
is the Minkowski metric with a signature chosen to be $(+,-,-,-)$,
and $i,j,k,...=0,1,2,3$.

However, more natural and favorable (for our consideration) is to write
down a density tensor of the angular momentum from fundamental
principles of the symmetry of electrodynamics under the Lorentz group
transformations i.e. by varying the electromagnetic field Lagrangian in
compliance with the Noether theorem (see, \hbox{for example,
\cite{ry})}.  In this case the angular momentum density tensor
 has the form
\begin{equation}\label{n}
M_{ij}^{(N)k}=[M_{ij}^{{}k}+{1\over
4\pi c}(x_{i}(\partial_{l}A_{j})F^{lk}-
x_{j}(\partial_{l}A_{i})F^{lk})]+{1\over
4\pi c}(-A_{i}F_j^k+A_{j}F_i^k),
\end{equation}
where the first two
terms correspond to the density of the orbital momentum and the last
two terms describe the spin momentum of the electromagnetic field. It
is just the manifest representation of the angular momentum as the sum
of the orbital and the spin part enables one, as we will see below, to
understand the physical meaning of Eq.(1).

These two definitions, \p{s} and \p{n}, of the angular momentum differ
by a total derivative
\begin{equation}\label{td}
M_{ij}^{(N)k}-M_{ij}^{
k}={1\over 4\pi c}\partial_{l}(x_{i}A_{j}F^{lk}-x_{j}A_{i}F^{lk}),
\end{equation}
and
since physical observables are spatial integrals of the conserved
current densities and under the condition that the fields tend quickly
enough to zero at infinity, the integrals of \p{s} and \p{n}
taken over the volume of 3-dimensional space  coincide. However, if the
fields tend to zero rather slow or spread to infinity (as, for example,
the plane waves), then taking into account boundary conditions becomes
essential for the computation of the integrals of \p{s} and \p{n}. In
overwhelming majority of cases this leads to interesting
physical observations concerning (global) properties of the physical
system. On the other hand the neglect of boundary conditions may result
in incorrect physical conclusions.

The simplest example is the
calculation of the projection of the angular
momentum on to the momentum of a plane electromagnetic wave.
 If for this calculation we use Eq.~\p{s},
 then the result will be
identically zero. Indeed, the Poynting vector of the plane wave has the
form
\begin{equation}\label{p}
P_{\alpha}=\int{d^{3}xT_{\alpha}^{0}(x)}={n_{\alpha}\over 8\pi c}
\int{d^{3}x({\bf E}^{2}+{\bf H}^{2})},
\end{equation}
where $n_\alpha$ is the unit  vector along the direction of wave
propagation, and $E_\alpha=F_{\alpha 0}$ and $H_{\alpha}={1\over
2}\varepsilon_{\alpha\beta\gamma}F^{\alpha\beta}$ are electric and
magnetic field strength, respectively, satisfying (for the plane wave)
the relation
\begin{equation}\label{he}
{\bf H}={\bf n}\times{\bf E}.
\end{equation}
The angular momentum calculated with the Eq.~\p{s} has the form
\begin{equation}\label{as}
M_{\alpha}=-{1\over
8\pi c}\varepsilon_{\alpha\beta\gamma}n_{\beta}\int{d^{3}xx_{\gamma}
({\bf E}^{2}+{\bf H}^{2})}, \end{equation} Thus, the helicity
\begin{equation}\label{sh}
s={{{\bf M}{\bf P}}\over|{\bf P}|}
\end{equation}
of any plane electromagnetic wave identically terns to zero if one
substitutes into \p{sh} Eq.~\p{p} and \p{as}, which runs counter the
observation: in general the plane wave has nonzero circular
polarization whose degree is  characterized just by the magnitude of
$s$ (or, another words, it is well-known that the photon has nonzero
helicity).

On the other hand the computation of $s$ with the use of
Eq.~\p{n} of the angular momentum density tensor gives the correct
result and allows one to find the connection between $h$ and $s$
helicity of the electromagnetic wave.

It is convenient to carry out the computation in the Coulomb gauge
\begin{equation}\label{cg}
A_{0}=0,\hskip18pt{\partial\over\partial{x^\alpha}}A^{\alpha}(x)=0,
\end{equation}
then the only contribution into the value of $s$ \p{sh} is given by the
spinning part of the angular momentum \p{n}, and Eq.~\p{sh} (with taking
into account \p{he}) takes the form
\begin{equation}\label{ara}
s=-{1\over 4\pi c}\varepsilon_{\alpha\beta\gamma}\int{d^{3}xn_{\alpha}
A_{\beta}E_{\gamma}}={1\over 4\pi c}\int{d^{3}xA_{\alpha}H_{\alpha}}=
{1\over 4\pi c}\int{d^{3}x{{\bf A}}(\nabla\times{\bf A})},
\end{equation}
which coincides with the $h$--helicity  (Eq.~(1)) of the
electromagnetic field.

Thus, the $s$--helicity  (or circular polarization) of the plane
electromagnetic wave is determined by Eq.~(1), which allows one to make
a conclusion, that the integral lines of electromagnetic field with
nonzero circular polarization have the nontrivial topological structure
characterized by a linking number.

\section{Angular momentum flow and
 $h$--helicity}

Let us show that the $h$--helicity  of an arbitrary electromagnetic
field is equal to the trace of the spin component of the angular
momentum flow.

It follows from the conservation law for the angular momentum \p{n}
\begin{equation}\label{cl}
\partial_{0}M^{(N)}_{\alpha\beta,0}-
\partial_{\gamma}M^{(N)}_{\alpha\beta,\gamma}=0
\end{equation}
that the density of the spin momentum flow has the form
\begin{equation}\label{g}
G_{\alpha\beta}=-{1\over
2\pi c}\varepsilon_{\alpha\gamma\delta}A_{\gamma}F_{\delta\beta},
\end{equation}
and
\begin{equation}\label{tr}
G\equiv TrG_{\alpha\beta}=-{1\over
2\pi c}\varepsilon_{\alpha\gamma\delta}A_{\gamma}F_{\delta\beta}=
{1\over 4\pi c}{{\bf A}}(\nabla\times{\bf A}).
\end{equation}
Thus, the density of the spin momentum flow along the direction of
electromagnetic field propagation (i.e. the density of $s$--helicity
flow) is determined by the $h$--helicity  density, and Eq.~(1)
characterizes the average value, over spatial volume, of the
$s$--helicity flow density. The topologically nontrivial
 configurations of the electromagnetic field manifest themselves through
the nonzero helicity values.

\section{Helicity in turbulence}

Let us now proceed with studying the relationship between the helicity
and angular momentum of turbulent flows. As we have already mentioned
in the Introduction the notion of the helicity of random fields plays an
important role in the theory of turbulence; thus, establishing the
connection between these two physical notions may occur to be useful
for better understanding the turbulence phenomenon.

We shall consider this problem by the example of the most simple type
of turbulent motion being homogeneous in space and time and isotropic
in space. Just that every case has been considered for studying
problems of magnetic field generation ($\alpha$--effect) \cite{m} and
the generation of large scale structures by small scale helical fields
\cite{mst}.

The pair correlation tensor of the velocity field  \hbox{${\bf V}({\bf
r},t)$}
\begin{equation}\label{4.1}
\left\langle V_i({\bf r},t)V_j({\bf
r^\prime},t^\prime)\right\rangle\equiv
B_{ij}({\bf r},{\bf r^\prime},t,t^\prime).
 \end{equation}
(`$\langle{~}\rangle$'denotes
statistical averaging) is the most
important characteristic of turbulence. So let us discuss this
characteristics in more detail.
The requirement for the tensor \p{4.1} to conserve the form under
  rotations and shifts in the space leads to the following  most
general form of the correlation tensor \cite{r}
\begin{equation}\label{4.2}
\left\langle
V_iV_j\right\rangle_{{\bf R},\tau}=A(R,\tau)\delta_{ij}+
B(R,\tau)R_iR_j+G(R,\tau)\varepsilon_{ijl}R_l,
\end{equation}
where  ${\bf R}={\bf r}-{\bf
r^\prime}$ and $\tau=t-t^\prime$, $A(R,\tau)$ and $B(R,\tau)$ are
scalar functions and $G(R,\tau)$ is a pseudoscalar function. All this
functions are expressed through the corresponding correlation functions
of \hbox{${\bf V}({\bf r},t)$}. For example,
$$
A(0,0)={1\over 3}\left\langle V_i({\bf r},t)V_i({\bf
r},t)\right\rangle
$$
has the physical meaning of the average energy density of the
fluid, and
\begin{equation}\label{4.3}
G(0,0)={1\over 6}\left\langle{\bf V}({\bf r},t)\bigl(\nabla\times{\bf
V}({\bf r},t)\bigl)\right\rangle
\end{equation}
determines the helicity density of {\bf V}.

The physical meaning of $B(R,\tau)$ is more
  complicated and we will not discuss it here.

One can easily see that the form
of the right hand side of \p{4.2} is conserved not only under the
spatial rotations and shifts but under the reflections as well. But
since the statement that turbulence is invariant under some group
transformations means that all correlation characteristics of
turbulence (and not only ,for example, \p{4.2}) are forminvariant under
this transformations, and \p{4.3} is indeed reflection noninvariant, it
follows that turbulence with nonzero $G(0,0)$  is reflection
noninvariant.

Let us show now that in agreement with the results of the previous
sections concerning the helicity of the electromagnetic field the
helicity of the turbulent velocity field also has (up to a
dimensional factor) the ``dynamical'' meaning of the projection of the
total orbital momentum onto the total momentum of the turbulent medium.

Assuming for simplicity that the density $\rho$ of the fluid is constant
we wright down the projection in the form
\begin{equation}\label{4.4}
s={\rho\over {\cal V}}\bigl\langle\int d{\bf r}[{\bf r}\times{\bf
V}({\bf r},t)] \int d{\bf r^\prime}{\bf V}({\bf
r^\prime},t)\bigr\rangle,
\end{equation} where ${\cal V}$ is the volume
of the medium.  Then, using \p{4.2}, the equation \p{4.4} may be
reduced to the form
\begin{equation}\label{4.5}
s=\rho\int d{\bf
R}R^2G(R,0)=\rho\int d{\bf R}R^2\left\langle{\bf V}({\bf
R},0)\nabla\times{\bf V}({\bf R},0)\right\rangle.
\end{equation}
If for $R$
larger then a correlation radius $G(R,0)$   tends to zero quick enough
we get from \p{4.5} that
\begin{equation}\label{4.6}
s\sim\left\langle{\bf V}({\bf r},t)\nabla\times{\bf V}({\bf
r},t)\right\rangle.
\end{equation}

Eqs.~\p{4.5}, \p{4.6} represent the relationship we have been
looking for. The observation of this fact points to the close
connection between topological properties and dynamical
characteristics of turbulence. Such a connection plays an important
role for understanding the processes of helicity (or angular
momentum) exchange between turbulent flows and external fields acting
on the former, the generation of various turbulence structures
possessing helicity, etc. \cite{m,ty,mst}.
 A simple example of helicity and angular
momentum transfer is considered in the next section.

Using the results obtained we may go even further and introduce the notion
of helicity for a system of particles by defining a microscopic
helicity as
\begin{equation}\label{4.7}
s^M=\sum_a[{\bf r}_a\times{\bf p}_a]\sum_b{\bf p}_b,
\end{equation}
where ${\bf r}_a$ and ${\bf p}_a$ are the radius vectors and the
momentums of the particles, $a,~b=1,...,N$ and $N$ is the number of
particles.  Using the microscopic phase density \cite{k}
\begin{equation}\label{4.8}
{\cal N}({\bf r},{\bf p},t)=\sum_a\delta\left ({\bf r}-{\bf r}(t)\right
)\delta\left ({\bf p}-{\bf p}(t)\right )
\end{equation}
we can rewrite \p{4.7} as
\begin{equation}\label{4.9}
s^M={1\over 2}\int d\Omega d\Omega^\prime[{\bf p}\times{\bf p}^\prime]
({\bf r}-{\bf r}^\prime){\cal N}({\bf r},{\bf p},t)
{\cal N}({\bf r}^\prime,{\bf p}^\prime,t),
\end{equation}
where $d\Omega$ is the measure of integration over the phase space.

Taking into account the fluctuation of the microscopic phase density
$$
\delta{\cal N}={\cal N}-\overline{\cal N},
$$
where $\overline{\cal N}=nf({\bf r},{\bf p},t)$, $f({\bf r},{\bf p},t)$
is the one
particle distribution function and $n={\cal N}/{\cal V}$ is the average
particle density in the system, we write down the average value of the
microscopic helicity as follows:
\begin{eqnarray}\label{4.10}
s=\left\langle s^M\right\rangle={1\over 2}\int
d\Omega d\Omega^\prime[{\bf p}\times{\bf p}^\prime] ({\bf r}-{\bf
r}^\prime)f({\bf r},{\bf p},t)f({\bf r}^\prime, {\bf
p}^\prime,t) \nn
+{1\over 2}\int d\Omega d\Omega^\prime[{\bf p} \times{\bf
p}^\prime]({\bf r}-{\bf r}^\prime)\left\langle\delta{\cal N}
({\bf r},{\bf p},t)\delta{\cal
N}({\bf r}^\prime,{\bf p}^\prime,t)\right\rangle.
\end{eqnarray}

The first term in \p{4.10} describes a ``hydrodynamic'' part of the
average helicity and can be rewritten in the form (compare with
\p{4.4})
\begin{equation}\label{4.11}
n^2 m^2\int d{\bf r}d{\bf r}^\prime[{\bf V}({\bf r},t)\times{\bf
V}({\bf r}^\prime,t)]({\bf r}-{\bf r^\prime}),
\end{equation}
where ${\bf V}({\bf r},t)={1\over m}\int d{\bf p}{\bf p}
f({\bf r},{\bf p},t)$ and $m$ is the particle mass.

The second term characterizes a ``correlation'' part of the average
helicity. If the correlation function of the phase density fluctuation
is isotropic the correlation part of the average helicity turns to
zero. One may also show that the correlation function of a weak
interacting homogeneous classical gas does not contribute, in the
 Landau approximation, into the helicity, and, hence, an inhomogeneous
state should be taken into account. Thus, for these cases the kinematic
definition of the helicity of a particle system
introduced above coincides with the hydrodynamic definition of the
helicity. In general case the correlation part of the helicity may
cause new correlation effects, but this point requires additional
studies.

\section{Angular momentum transfer from field to medium}

In previous sections we have elucidated the connection between the
angular momentum and the $h$--helicity of the electromagnetic field and
medium.

In many physical applications a problem arises to determine the value
of energy, momentum and angular momentum transferred from the  field to
the medium. This problem is important for understanding and predicting
the behavior of the system.

However, while, for example,  in plasmas
and turbulence the energy and momentum transfer have been studied in
detail \cite{stur}, the problem of angular momentum transfer has not
been developed in the same extent, though it has been under
consideration since Sadovski found the effect of turning a small plate
by an electromagnetic wave \cite{sad}.

We see that the  problem of angular
momentum transfer is related to the problem of the helicity transfer
from fields to the medium discussed, for example, in \cite{t}.

 The problem of angular momentum
transfer from electromagnetic fields to particles and consequences of
this effect for the electromagnetic pumping up of the angular momentum
into plasmas were discussed, in particular, in Ref.~\cite{so}.   Here
we consider this question from somewhat different point of view based
on the relationship between the angular momentum and the $h$--helicity,
and, in contrast to \cite{so}, demonstrate that the charged
particles may acquire orbital momentum from {\it plane }
electromagnetic waves with nonzero helicity and that this effect may
result in pumping up the angular momentum into plasma.

We restrict ourselves by the consideration of the simplest possible
model of a system of non-interacting charged particles, so that the
problem is actually reduced to considering the motion of a single
particle.

Let a particle with a charge $e$ and a mass $m$ be placed into the
field of a circular polarized plane wave propagating along the axis $x$
  and having electric strength
\begin{equation}\label{es}
{\bf E}={\bf e}_yE_{0}\cos{\omega(t-{x\over c})}+\alpha{\bf e}_zE_{0}
\sin{\omega(t-{x\over c})},
\end{equation}
where $E_0$ is the wave amplitude, $\omega$ is the electromagnetic field
frequency and $\alpha=\pm 1$ denotes   left or right polarization.

Note that due to Eqs.~\p{sh}, \p{ara} the angular momentum of the plane
wave is directed along the wave momentum (axis $x$), and
its value coincides with the helicity of the electromagnetic wave,
the helicity density of the latter being
\begin{equation}\label{hd}
G={\alpha\over 4\pi c}{{\bf A}}\nabla\times
{{\bf A}}={\alpha E_{0}^{2}\over
4\pi\omega}={\alpha\varepsilon\over\omega},
\end{equation}
where $\varepsilon={E_{0}^{2}\over 4\pi}$ is the energy density of the
wave. If we considered a wave with arbitrary polarization the value of
$\alpha$ in \p{hd} would have  determined the corresponding
polarization properties of the wave.  For instance, if
the  wave is linear polarized, $\alpha=0$, and, hence $h=s=0$, which
testifies to the triviality of the corresponding topological
characteristics of the linear polarized wave.

The equation of motion of the particle in the electromagnetic wave is
nonlinear, and, in general, finding an exact solution of the equation is
nontrivial problem.\footnote{Note that the well known particular
solution of the equation, which describes particle motion along a
circle perpendicular to the wave momentum, implies that the particle
 had the angular momentum {\it a priori}, and the latter does not vary
in the course of motion. Hence, this solution is not one we are looking
for.} In view of this, and since here we do not aim at carrying out the
rigorous derivation of the expression for the particle orbital momentum
acquired, we shall consider the process of angular momentum transfer
from the wave to the particle using semiqualitative consideration
analogous to that used, for example, in the computation of the strength
of high frequency pressure \cite{chen}.

We look for the solution as a series expansion in the field amplitude
and restrict ourselves to the nonrelativistic  approximation. Particle
motion consists in quick rotation in $(y,z)$ plane and slow motion
averaged over the oscillations. Let us further assume that the
particle velocity $\langle{\bf v}\rangle$ averaged over the
oscillations is directed along wave propagation (axis $x$). We shall
see below that for particle to get an angular momentum from the
wave it is crucial that the former accelerates along the wave.
Slow variation of $\langle{\bf v}\rangle$ with time may be caused, in
the first place, by the momentum transfer from the scattered wave to
the particle, or by a weak constant electric field.

Then from the equation
\begin{equation}\label{.4}
m{{d^2{\mathaccent "7E {\bf r}}}\over{dt^2}}=e{\bf E}(\langle{\bf
r}\rangle ,t)+{e\over c}\langle{\bf v}\rangle\times {\bf H}(\langle{\bf
r}\rangle ,t)
 \end{equation}
for the oscillating part ${\mathaccent "7E
{\bf r}}$ of the particle radius vector we get
\begin{equation}\label{.5}
{\mathaccent "7E {\bf r}}=-{{eE_0}\over{m\omega^2(1-{\langle{v}\rangle\over
c})}}\bigr({\bf
e}_yE_{0}\cos{(1-{\langle{v}\rangle\over c})\omega t}+ \alpha{\bf e}_zE_{0}
\sin{(1-{\langle{v}\rangle\over c})\omega t}\bigl).
\end{equation}
$(\langle{v}\rangle\equiv |\langle{\bf v}\rangle |)$

Eq.~\p{.5} may be directly obtained by solving Eq.~\p{.4} in
the coordinate system where $\langle{\bf v}\rangle$ is equal to zero in a given
time moment and by passing to the laboratory system with making use
of the Lorentz transformation (Doppler effect) in the nonrelativistic
approximation.

Momentum variation with time is expressed by the formula
\begin{equation}\label{.6}
{d{\bf M}\over{dt}}=\bigl[{\mathaccent "7E {\bf r}}\times
m{{d^2{\mathaccent "7E {\bf r}}}\over{dt^2}}\bigr]
\end{equation}

Taking into account slow dependence of $\langle{\bf v}\rangle$ on time
we get from \p{.5} and \p{.6} that
\begin{equation}\label{.7}
{d{\bf M}\over{dt}}={{4\pi e^2}\over{mc\omega^2}}{{d\langle{\bf
v}\rangle}\over{dt}}G
 \end{equation}

We see that the rate of momentum variation is proportional to the
helicity density $G$ and particle's slow acceleration along the wave.
This testifies to the existence of the effect of momentum transfer from
an electromagnetic wave to a charged particle, and indicates that such
a transfer is possible if only the wave possesses the helicity, the
process being governed by the angular momentum conservation law. The
simple example having been considered shows a way the topologically
nontrivial fields affect the dynamics of the charged particles.

For plasmas this effect means the possibility of pumping up the angular
momentum into the plasmas by the helical fields. The angular momentum
of the plasma will increase until the particle acceleration along the
wave is compensated by the friction force, and until the increase is
terminated because of the angular momentum loss caused by particle
radiation.

\section{Conclusion}

For electromagnetic fields, turbulent media and particle systems we
have established the relationship between the $h$--helicity (1), being
a characteristics of topologically nontrivial vector fields,
and the s--helicity \p{sh} which characterizes the value of the
projection of the angular momentum onto the momentum of the medium.
This testifies to the close connection between corresponding
topological and dynamical properties of the physical objects, which
should be taken into account when studying the effects of angular
momentum and helicity exchange between fields and media.

For hydrodynamics and turbulence the momentum transfer between fields
and media points to the ``intertwining'' of the helicity of the field
with that of the medium velocity field, which may point to the
existence of a conservation law for a topological charge characterizing
the field-medium system as a whole.

{\bf Note added.}
When this work was completed the authors became aware of Ref.~\cite{ra},
where the connection between the two types of helicity of the
electromagnetic field was pointed out in the framework of a
topological electromagnetism.

\bigskip
{\bf Acknowledgments}

\medskip
D.S. is grateful to the European Community for financial support under
the contract CEE-SCI-CT92-0789, the University of Padova and INFN
(Sezione di Padova), and in particular Prof.~M.~Tonin and Prof.
P.~A.~Marchetti for kind hospitality and valuable discussion.

\smallskip
The authors are grateful to Prof. A. I. Akhiezer and Prof. D. V. Volkov
for useful discussion.

\medskip
This work was supported in part by the Ukrainian State Committee in
Science and Technologies.

\end{document}